# Charge reconstruction study of the DAMPE Silicon-Tungsten Tracker with ion beams

Rui Qiao[1;], Wen-Xi Peng[1;1)], Dong-Ya Guo[1], Hao Zhao[1], Huan-Yu Wang[1], Ke Gong[1], Fei Zhang[1],
Xin Wu[2], Phillip Azzarello[2], Andrii Tykhonov[2], Ruslan Asfandiyarov[2], Valentina Gallo[2],
Giovanni Ambrosi[3], Nicola Mazziotta[4], Ivan De Mitri[5,6],

[1] Institute of High Energy Physics, Chinese Academy of Sciences, Beijing 100049, China
[2] Département de Physique Nucléaire et Corpusculaire, University of Geneva, Geneva, Switzerland
[3] Dipartimento di Fisica e Geologia, Università di Perugia, Perugia, Italy
[4] Istituto Nazionale di Fisica Nucleare Sezione di Bari, Bari, Italy
[5] Dipartimento di Matematicae Fisica "E. DeGiorgi", Università del Salento, Lecce, Italy
[6] Istituto Nazionale di Fisica Nucleare Sezionedi Lecce, Lecce, Italy

**Abstract**: The DArk Matter Particle Explorer (DAMPE) is one of the four satellites within Strategic Pioneer Research Program in Space Science of the Chinese Academy of Science (CAS). DAMPE can detect electrons, photons in a wide energy range (5 GeV to 10 TeV) and ions up to iron (100GeV to 100 TeV). Silicon-Tungsten Tracker (STK) is one of the four subdetectors in DAMPE, providing photon-electron conversion, track reconstruction and charge identification for ions. Ion beam test was carried out in CERN with 60GeV/u Lead primary beams. Charge reconstruction and charge resolution of STK detectors were investigated.

## 1 Introduction

The DArk Matter Particle Explorer (DAMPE) has been successfully launched on December 17, 2015. DAMPE can measure electrons and photons in 5GeV ~ 10 TeV, and ions in 100 GeV ~ 100 TeV up to Iron [1].

The layout of DAMPE is shown in Fig. 1. DAMPE is made up of 4 subdetectors: plastic scintillator detectors (PSD), a silicon-tungsten tracker-converter (STK), a Bismuth Germanium Oxide imaging calorimeter (BGO) and neutron detectors (NUD). The PSD consists of 2 layers of scintillators to provide electron/gamma separation and charge identification for ions up to Iron. The STK consists of 12 layers of single-sided silicon detectors with a total active area of 6.6 m$^2$. STK can provide tracks reconstruction and charge identification for the incoming charged particles. The expected charge resolution of STK is better than 10% for electrons and angular resolution better than 0.2 degrees. The BGO calorimeter is made of 14 layers of BGO bars, with a total thickness equivalent to 31 radiation lengths and 1.6 nuclear interaction lengths. The expected energy resolution of BGO is better than 1.5% for 800GeV electrons and 40% for 800GeV nucleus. BGO also provide triggers for the other 3 payloads. The NUD consists of 4 boron-doped plastic scintillators with size of 19.5*19.5*1 cm$^3$. The NUD can detect neutrons of showers from electrons and protons, which may improve the electron/proton separation power.

* Supported by National Natural Science Foundation of China (11403025), State's Key Project of Research and Development Plan (2016YFA0400204), Strategic Pioneer Research Program in Space Science of the Chinese Academy of Science
1) E-mail: pengwx@ihep.ac.cn, qiaorui@ihep.ac.cn

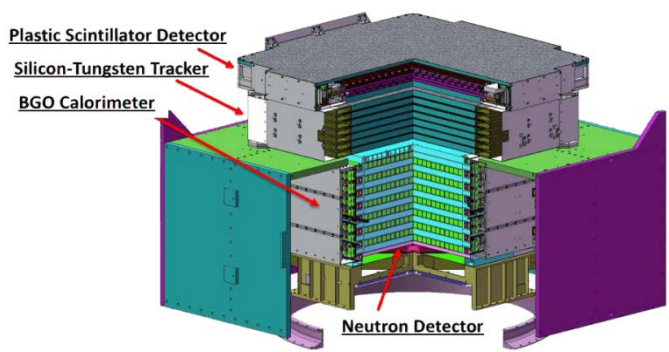

Fig. 1. Schematics of the DAMPE detector.

The STK consists of single-sided silicon microstrip detectors and tungsten plates, which is similar to the trackers of FERMI [2] and AGILE [3]. A detail description of STK detector can be found in ref 4. The tungsten plates convert incident photons into $e^-$ and $e^+$ pairs, and the silicon detectors will track the resulting $e^-$ and $e^+$ pairs as well as the other charged particles. What's more, the STK is also designed to identify ions up to iron ($Z$=26).

There are 6 planes of dual-orthogonal layers of AC-coupled single-sided silicon microstrip detectors, with a total active area of 6.6 $m^2$. Each layer has 64 silicon microstrip detectors with a dimension of $95 \times 95 \times 0.32$ $mm^3$ each. Each detector consists of 768 p+ implanted strips with a pitch of 121 µm and width of 48 µm. Every 4 detectors are grouped together, named ladder, sharing the same bias supply and front-end electronics. This helps to reduce the power consumption and the number of electronics, and has been proved useful by previous space experiments [2, 3, 5]. Furthermore, only half of 768 strips are connected to the front-end electronics. These 384 strips are named readout-strips while the others are named float-strips.

The signal from 384 readout-strips are shaped and amplified by 6 VA140 ASIC [6] integrated on the ladder. The VA140 is a 64 channel charge sensitive preamplifier-shaper circuit with a low noise level (425$e^-$ for 50pF) and low power consumption (0.29mW/Channel). The analogical VA140 output signals are converted into differential signals first, and then transferred into Tracker Readout Board (TRB) [7] where the signals are digitized and readout. There are several data processing modes for TRB, and the most commonly used are:

- Data compression mode, where pedestal subtraction and cluster finding [8] are applied to all signals. Only the amplitudes of channels inside the found clusters will be output, while the information of other channels will be abandoned. This help to compress output data size by a factor of 2.
- Pedestal update mode, where TRB automatically calculate the pedestal of each channels using 1024 periodical triggers (50Hz in this paper). The calculated pedestals will be stored and used by data compression mode.

Tungsten plates are both 1mm thick and assembled on the top of plane 2, 3, 4 of silicon detectors. The plane 1 of silicon detectors can provide accurate position measurement of incident charged particles without multi-scattering by tungsten plates.

## 2 Experiment setup

The ion beam test was carried out at CERN SPS (Super Proton synchrotron) in November 2015 in order to investigate the charge measurement behavior of STK ladders. The primary beams were 60 GeV/n Lead and secondary fragments were selected with $A/Z$ = 2. 8 flight model STK ladders in parallel direction were installed perpendicular to the beam. 2 plastic scintillators readout by SiPM were installed upstream and downstream of STK ladders separately, providing charge identification for incident fragments. Unfortunately, the downstream plastic scintillator suffered from poor charge resolution and could not be used for charge identification. The layout of beam test is shown in Fig. 2.

To study multiple-scattering from tungsten converter and support structure, 12 orthogonal single-sided silicon microstrip detectors (SSDs) were installed between plastic scintillator and STK ladders. As the capacitances of these SSDs were different from STK ladders, and multiple-scattering is of no interest in this paper, these SSDs will not be analyzed and discussed.

Both STK ladders and SSDs were readout and digitized by two flight model Tracker Readout Board (TRB), which worked in data compression mode and pedestal update mode alternatively. As the accuracy of pedestals were critical to the accuracy of cluster amplitudes, pedestal update mode were carried out every 4 hours to eliminate pedestal fluctuation.

At the end of experiment, the 8 STK ladders were inclined to 9 and 10 degrees for a short period. But their statistics were so limited, so only charge reconstruction using perpendicular incident events is discussed in this paper.

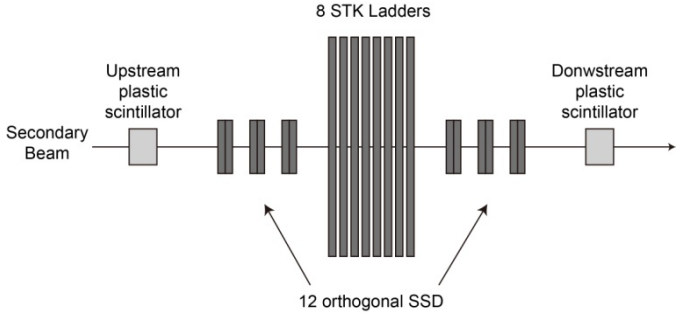

Fig. 2. Layout of beam test.

## 3  Data Analysis

Data analysis was divided into four sections:
- Charge identification by plastic scintillators.
- Data selection of STK ladders.
- Electronics saturation.
- Charge reconstruction and resolution.

### 3.1  Charge identification by plastic scintillators

The interaction between highly energetic charged particles and plastic scintillator is mainly through ionization. The deposited energy, which is proportional to $Z^2$ according to Bethe-Bloch formula [9], can be used to identify the charge of incoming particle. Although the deposited energy is also relative to other parameters, such as velocity and path length, but these parameters were fixed during the beam test. So we simply took the square root of deposited energy as the estimator of charge $Z$ for beam test data.

To be mentioned, the light output of plastic scintillator is almost proportional to the deposited energy for light ions, but become slowly saturated for heavy ions, which is also known as Birk's saturation [10]. So it was hard to identify ions heavier than Titanium ($Z$=22) by the plastic scintillator used in this experiment. Although Birk's saturation can be corrected [11], the resolution of charge identification in this experiment cannot be improved. Finally, we simply took the square root of integration of the SiPM signals as the estimator of charge $Z$ for incident fragments.

As the gain of SiPM is sensitive to the bias voltage and temperature which might change during experiment, gain equalization was carried out for each data file. The mean value of charge peaks were used to evaluate the gain of SiPM during each data file. The Gaussian mean value of charge peaks could be achieved by multi-Gaussian fits to the charge spectra, but the results might be ambiguous if the statistic was low. A better way was to generate single ion charge spectrum and fitted with single Gaussian distribution. The single ion charge spectrum of upstream plastic scintillator was accumulated with cuts on the downstream plastic scintillator amplitude.

After gain equalization, the charge spectra of upstream plastic scintillator from different data files could be merged, as shown in Fig. 3. Multi-Gaussian fit results and the corresponding charge numbers are also shown in Fig. 3. The charge peaks of light ions were quite clear, while the charge peaks of heavy ions were harder to distinguish as a result of Birk's saturation. In principle, light ions were dominated in secondary fragments, but the most dominant ion in Fig. 3 is Oxygen. This is because high trigger threshold was applied in this experiment to increase the trigger rates for heavy ions.

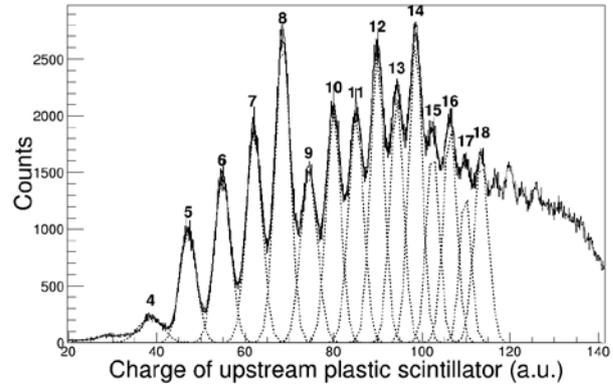

Fig. 3. Charge spectrum of upstream plastic scintillator

### 3.2  Data selection of STK ladders.

#### 3.2.1  Light ion removal

Before removal, each STK ladders collected around 15 clusters in each event. As we applied a 5$\sigma$ threshold for cluster findings, these 15 clusters were attributed to 15 incident fragments. These fragments can be divided into 3 types:

- Ions (heavy than Lithium) identified by upstream plastic scintillator.
- Ions out of the integration window of SiPM and could not be identified by upstream plastic scintillator. On the other hand, STK ladders had a much longer integration window (6.5 μs) and could still collect these signals.
- Light ions inside the integration window of SiPM but contributed only a little deposited energy. As mentioned above, the deposited energy is proportional to $Z^2$. That means the contribution from light ion is much less than that from heavy ion.

The ions of the last 2 types had different charges from that identified by upstream plastic scintillator, and should be excluded before further analysis. We sorted all the clusters in each STK ladders according to the total cluster amplitudes. Although STK charge reconstruction suffered from saturation and position dependency as described below, most clusters with the second largest total cluster amplitudes could be assigned to Helium and Lithium. That means cluster with the largest total cluster amplitude was the ion identified by upstream plastic scintillators. The other clusters were mostly light ions and excluded.

### 3.2.2 Alignment and perpendicular incident event selection

As the charge reconstruction of STK ladders is also relative to the incident angle, inclined incident events should be removed from the dominant perpendicular incident events. First of all, alignment of 8 STK ladders was carried out according to the center-of-gravity (COG) of clusters on 8 ladders. The difference of cluster COG from the 2 most distinct ladders is shown in Fig. 4. The sigma value of Gaussian fit is 0.13 mm. Then $4\sigma$ cuts were applied to the difference of cluster COG to exclude inclined incident events. As the beam illuminated mostly on a beam spot of diameter 20 mm, events outside the beam spot were also excluded.

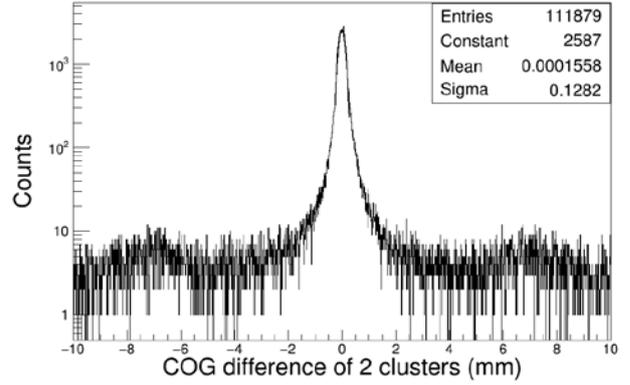

Fig. 4. Distribution of cluster COG difference from the 2 most distinct STK ladders.

### 3.2.3 Readout-strip and float-strip incident event selection and gain equalization

Similar to plastic scintillator, the square root of the total cluster amplitudes can be an estimator of the charge of corresponding fragment. Although silicon microstrip detectors do not suffer from Birk's saturation, they suffer from position dependency [3, 12]. For STK ladders, the readout-strip incident events have the largest total cluster amplitudes and the float-strip incident events have the smallest total cluster amplitudes. Ion hit on the intermediate region results in position-relative medium total cluster amplitudes.

The portion of readout-strip, float-strip and intermediate region incident events depend on the incident angle. The portion of intermediate region incident events grows for inclined incident events, while the readout-strip and float-strip events dominate in perpendicular incident events. With perpendicular incident events selected by previous step, the charge spectrum of STK ladders is dual-Gaussian like [12], mostly contributed by the readout-strip and float-strip incident events separately. Separation of readout-strip and float-strip incident events could help to greatly improve the charge resolution.

The COG of cluster can be a potential estimator to identify readout-strip and float-strip incident events. Clusters of readout-strip incident event have a COG close to the readout-strip, and vice versa. However, the COG may be biased by the low signal-to-noise ratio channels in the cluster and the resolution is not good enough. A better estimator was the COG of only 2 channels: the channel with the largest amplitude (named seed channel) and its larger neighboring channel. This estimator was also named as $\eta$. The distribution of $\eta$ for Carbon ($Z=6$) is shown in Fig. 5a. The peaks close to 0 and 1 correspond to readout-strip incident events, while the peak close to 0.5 corresponds to float-strip incident events. Readout-strip incident events were defined by $\eta \in [0,0.1]$ and $[0.9,1]$, while the float-strip incident events were defined by $\eta \in [0.4,0.6]$. The dual-Gaussian like charge spectrum of Carbon is shown in Fig. 5b. The contributions by readout-strip and float-strip incident events are shown in blue and red spectrum, respectively.

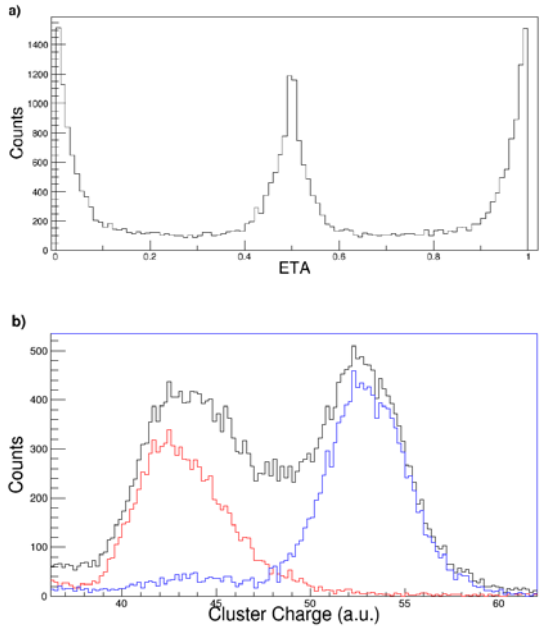

Fig. 5. Distribution of $\eta$ (a) and the charge spectrum of Carbon (b).

Compared to SiPM, the gains of STK detectors were quite stable because the detectors were always over depleted. The variation of gain might be mostly attributed to the successive electronics. As the basic element of successive electronics is VA140 (64 channels), we accumulated charge spectra based on VA140. By comparing the peak position from the first and the last data files, the gain fluctuations of electronics were found to be smaller than 0.5% in average. As a result, we assumed that the gain of STK ladders were stable during the whole experiment, and the charge spectra from the same VA140 in different data files could be merged together to increase statistics.

The peak positions of charge spectra from either readout-strip or float-strip incident events could be used to process gain equalization. Finally, we used the mean value of Oxygen ($Z=8$) charge spectra from float-strip incident events to process gain equalization for different VA140s and successive electronics. Charge spectra of lighter ions were not preferred as they suffered from low statistics, as shown in Fig. 3. Charge spectra from heavier ions or from readout-strip incident events were also not preferred, as they suffered from saturation discussed below.

### 3.3 Electronics saturation

After data selection and gain equalization, the charge spectra from both readout-strip and float-strip could be finally achieved, separately. But the square root of the total cluster amplitudes became 'saturated' for heavy ions from both readout-strip and float-strip incident events. Detail analysis showed that not all channels in a heavy ion cluster suffered from saturation, but only those channels with amplitudes larger than a threshold around 3000 ADC counts. The square root of seed channel amplitudes as a function of identified charge from both readout-strip and float-strip incident events are shown in Fig. 6. For readout-strip incident events, the response seems linear up to Carbon ($Z=6$), and then become saturated. The charge response of float-strip incident events is similar, except the linearity range increases to Neon ($Z=10$). As the turning points of both readout-strip and float-strip are around 55 (corresponding to 3000 ADC counts) in Fig. 6, this saturation effect might be attributed to the electronics. This was confirmed by the result of Tracker Readout Board (TRB) running at calibration mode. During the calibration mode of TRB, several tens of distinct calibrated charges were injected to the calibration inputs of VA140 and the corresponding amplitudes of each channel were digitalized. Each calibrated charge was generated via a programmed controlled DAC and a 2 pF capacitance. If the electronics were linear, the channel amplitudes should be proportional to the controlled DAC voltage. However, the calibration results confirmed that the electronics of STK became greatly saturated from around 3000 ADC counts.

As the results of TRB calibration mode were not reliable enough to do a saturation correction, we excluded the saturation events which defined by the amplitude of seed channel larger than 3000 ADC counts. Readout-strip incident events were more likely to be excluded than float-strip incident events, as readout-strip incident events had much larger seed channel amplitudes. The efficiency of Boron (Z=5) was 100% with this cut, while the efficiencies became lower for heavier ions. Charge identification for ions heavier than Neon (Z=10) is impossible in this paper.

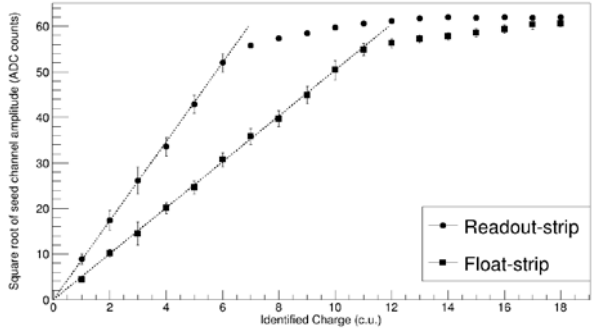

Fig. 6. The correlation between the seed channel amplitudes and identified charge

### 3.4 Charge reconstruction and resolution

Silicon microstrip detectors are periodical structure, as long as they are perfectly manufactured. The basic periodical element is the region from the center of a readout-strip to the center of his neighboring readout-strip. The incident ion position within this region can be characterized by $\eta$ defined above.

As mentioned above, readout-strip incident events have the largest total cluster amplitudes while the float-strip incident events have the smallest amplitudes. This position dependency was corrected according to the charge spectra with binning on $\eta$. The bin width was chosen to be 0.05, which was a compromise between accurate correction results and sufficient statistic for fitting. The correlations between the fitted mean values of charge spectra and $\eta$ from Boron (Z=5) to Oxygen (Z=8) are shown in Fig. 7. The fitted mean values were normalized for easy comparison. Only $\eta$ from 0 to 0.5 is shown as a result of symmetry. Fig. 7 shows that the correlations from Boron to Oxygen are the same within the range of error. This is because the position dependency is mostly attributed to the capacitances of silicon microstrip detector [13]. As a result, we used the weighted mean of correction factors from Boron to Oxygen to be the final correction factors. To be mentioned, Nitrogen and Oxygen have no fitted mean values for $\eta$ close to 0 in Fig. 7, as these events suffered from electronics saturation and excluded.

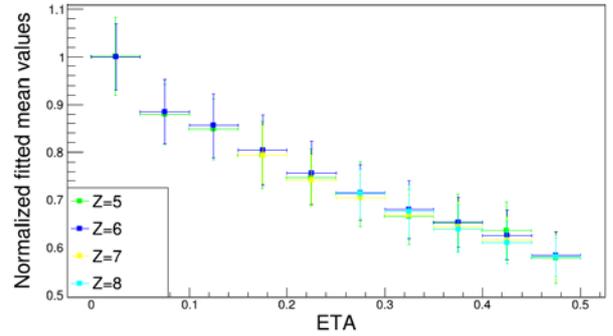

Fig. 7. The normalized fitted mean values of position-dependent charge spectra from Boron (Z=5) to Oxygen (Z=8).

After position dependency correction, the charges of STK ladders were finally reconstructed. The charge resolutions were better than 0.3 change unit from Hydrogen to Neon, as shown in Fig. 8. These resolutions were comparable to those of AMS-02 [12]. Although the correction factors were generated according to ions from Boron to Oxygen, they seem also applicable for other ions. Hydrogen and Helium still exist in the charge spectra after light ion exclusion. These might be attributed to ions break-up when interacting with tungsten converter and support structure. As these events fulfilled the perpendicular incident selection, their charges were also well reconstructed. Lithium is absent in Fig. 8 as the statistics were too limited to do fitting.

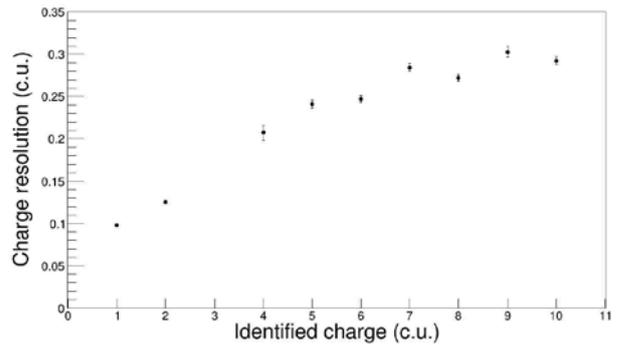

Fig. 8. Charge resolution.

## 4 Conclusion

The Silicon-Tungsten Tracker (STK) was designed to provide photon-electron conversion, track reconstruction and charge identification for ions up to Iron. The charge reconstruction and resolution of STK ladders were studied using $A/Z=2$ fragments from 60GeV/n Lead primary beams at CERN SPS.

It was found that STK suffered from electronics saturation for channel amplitudes larger than around 3000 ADC counts. Ions lighter than Carbon ($Z=6$) were not affected by this saturation while ions heavier than Neon ($Z=10$) were fully affected. Ions from Carbon to Neon were partly affected, depend on the incident position. Readout-strip incident events were more likely to suffer from saturation while the float-strip incident events were less affected. We used a 3000 ADC counts threshold to exclude saturation events in this paper.

The position dependency of STK was studied using ions from Boron ($Z=5$) to Oxygen ($Z=8$). These 4 position dependency correlation agreed with each other within the range of error. The position dependency correction factors were applied to reconstruct ion charges from Hydrogen to Neon. The charge resolutions were better than 0.3 charge unit.